# The variant of post-Newtonian mechanics with generalized fractional derivatives

V. V. Kobelev[*]

**Abstract**

In this article we investigate mathematically the variant of post-Newtonian mechanics using generalized fractional derivatives. The relativistic-covariant generalization of the classical equations for gravitational field is studied. The equations match to (i) the weak Newtonian limit on the moderate scales, (ii) deliver a potential higher, that Newtonian, on certain, large-distance characteristic scales. The perturbation of the gravitational field results in the tiny secular perihelion shift and exhibits some unusual effects on large scales.

The general representation of the solution for fractional wave equation is given in form of retarded potentials. The solutions for the Riesz wave equation and classical wave equation are clearly distinctive in an important sense. The Riesz wave demonstrates the space diffusion of gravitational wave at the scales of metric constant. The diffusion leads to the blur of the peak and disruption of the sharp wave front. This contrasts with the solution of D'Alembert classical wave equation, which obeys the Huygens principle and does not diffuse.

---

[*] University of Siegen, PB A203, D-57076, Siegen, Germany
Manuscript submitted: 1/20/2011

V. V. KOBELEV

**Content**

THE VARIANT OF POST-NEWTONIAN MECHANICS WITH GENERALIZED FRACTIONAL DERIVATIVES

# 1 Introduction

The astronomic date witness much higher levels of gravitation potentials on the galactic scales, as those predicted by Newton law. The main motive of the purely phenomenological theory, proposed in this article, is the attempt to fit high-precision data of Solar system experiments to observable astronomic phenomena on galactic scales. Namely, we try to explain the distribution of gravitational potential, which follows from the solution of inverse problem from galaxy dynamics, without implication of dark matter. The hypothesis explains maximal sizes of galaxies and predicts the belts of stability in the outer periphery of the Galaxy, inhabited by dwarf galaxies, and matter-free belts of instability. This appearance proved to be conforming to the spatial distributions of neighboring members of Local Group. The observed peculiarities of the gravitational lensing by galaxies are also discussed.

A variety of high precision null experiments verify the weak equivalence principle and local Lorentz invariance, while gravitational redshift and other clock experiments support local position invariance. Together these results conform the Einstein Equivalence Principle, which underlies the concept that gravitation is synonymous to spacetime geometry, and must be described by a metric theory. Einstein general relativity postulates isotropic partial field equation of the order second. Moreover, the general relativity matches the Newton law as its the weak-field and non-relativistic limit. Any possible violation of the Newton law immediately influences the fundamental, yet postulated, linear relation between curvature and mass-energy tensors. Einstein's relativistic triumph of 1905 and its follow-up in 1915 altered the course of science. They were triumphs of the imagination and of theory; experiment played a secondary role. In the past four decades, we have witnessed a second triumph for Einstein, in the systematic, high-precision experimental verification of his theories [**1**,**2**].

The confirmation of the Newton law and, consequently, General relativity on the galactic scales is by no means perfect. The main problem is the existence of "dark matter" is inferred from astrophysical observations that probe gravitational potentials. The mass content required to provide the derived gravitational potential is then compared with the visible mass content. Some of the evidence is summarized below [**3**]:

1. While measuring rotational speeds of galaxies, astronomers perceived a systematic anomaly [**4**]. Galaxies seem to turn as if they constituted a quasy single block of matter. In other words, rotational speeds are too rapid at the periphery of galaxies, as compared to what they should be according to the Newtonian gravitation laws. In classical gravitation theory, this suggests the existence of an invisible heavy matter, the black matter, distributed in a halo surrounding galaxies. The dark matter must represent more than 90% of the total mass of the galaxy. Rotation curves for a large number of spiral galaxies have now been reliably established and it is observed that the orbital velocities of objects (stars, globular clusters, gas clouds, etc.) tend to a constant value, independent of the radial position, even for objects out toward, and even far beyond, the edge of the visible disks. This is quite inconsistent with the behaviour expected from Newtonian mechanics, assuming most mass is in the central part of the galaxies.
2. Studies of the dynamics of stars in the local disk environment gave rise to the first suggestion of dark matter. The kinetic energy associated with the motion of these stars normal to the plane of the Milky Way gives a measure of the restraining gravitational potential that binds them to the disk.

3. Within the Local Group, the Milky Way and Andromeda (M31) are approaching each other at a much faster pace than can be explained by gravitational attraction of the visible mass. To explain the approach velocity, and indeed the fact that these two galaxies are not still moving away from each other as part of the Hubble expansion, requires each to have masses that are consistent with those deduced from their rotation curves.
4. Many clusters of galaxies show extended x-ray emission. This is usually attributed to a thin plasma of hot gas. The common assumption is that the hot gas is gravitationally bound to the cluster and in equilibrium. The gravitational potential energy for the virial system can be inferred from the kinetic energy balance of the hot gas. The cluster mass determined in this way is much higher than that seen either visibly or in the gas itself.
5. Gravitational lensing by clusters of galaxies causes images of more distant galaxies to be distorted and often split into multiple images. The gravitational mass of the lensing cluster, and its distribution, can be recovered through detailed analysis of the image pattern surrounding the cluster. The lenses show a far more extended spatial extent than the visible cluster.
6. Galaxy red-shift surveys have revealed large-scale galaxy-cluster streaming motions superimposed on the Hubble expansion. Attempts to explain this phenomena due to gravitational attraction resulting from the overall distribution of galaxy superclusters give the right direction of motion but need more than the observed visible masses in the superclusters to explain the speed of motion.

In most cases the mismatch between the required mass and the observed mass is tremendous. The alternative interpretation of astronomic date presumes higher levels of gravitation potentials on the galactic scales, as those predicted by Newton law. The main reason of the purely phenomenological theory, proposed in this article, consist in the attempt to fit high-precision data of Solar system experiments to observable astronomic phenomena, namely, high level of gravitational potential, which follows from the a analysis of galaxy dynamics. We propose the relativistic-covariant generalization of the classical equations for gravitational field, which matches to (i) the weak Newtonian limit on the moderate scales, (ii) delivers a potential higher, that Newtonian, on certain, large-distance characteristic scales, abandoning the dark mass hypothesis. On the moderate scales, the strong field limit for the proposed model corresponds those of the classical General Relativity.

The present work continues the development of post-Newtonian mechanics using generalized fractional derivatives [**5**, **6**]. The model proposed is of pure speculative nature, but delivers an unusual explanation of some really observed phenomena. The central concept of our model consists of the introduction of isotropic partial field equation of the integer order lower or equal than second, i.e. of the polynomial of the second order. The classical General Relativity equations contain solely quadratic term, because there is no linear isotropic term of the first order in ordinary differential calculus. The corresponding spherically symmetric and Lorenz invariant Hamiltonian must contain fractional derivatives. We investigate the relativistic-covariant generalization of the classical equations for gravitational field, which matches to (i) the weak Newtonian limit on the moderate scales, (ii) delivers a potential higher, that Newtonian, on certain, large-distance characteristic scales. The governing equations are presumed to be generally covariant, relativistic invariant and spatially isotropic. This assumption leaves sufficient room for the generalization of the equations and explanation of the observation data.

We proceed with the idea of additional Riesz potential term [**7**]. This term reveals scale-dependent gravity with a sole metric parameter $\lambda$ with dimension of length. The character length $\lambda$ is approximately the distance from the Sun to Galaxy center. The Riesz potential matches the Newtonian potential on the short scales. The perturbation of Keplerian laws Solar system is tiny, but observable. The perturbation of the gravitational field results in the tiny secular perihelion shift. The secular perihelion shift increases with the size of the system. On galactic sizes the Riesz potential declines slowly then Newtonian and the corresponding acceleration is roughly three times higher that Newtonian. The high rotational speed of Sun and other peripheral stars is the direct consequence of stronger than Newtonian gravitational acceleration towards the center of Galaxy.

## 2 Mathematical preliminaries: tensor analysis with fractional derivatives

### 2.1 Semi-covariant and semi-contravariant fractional derivatives

The formal description of hypothesis is essentially based on a special kind of pseudodifferential equations. For the formulation of the field equations we assemble the necessary mathematical apparatus.

The procedure is based on the Riemann-Liouville approach to the fractional differentiation of scalar function $f = f(t), -\infty < t < \infty$. The left-hand and right-hand Riemann-Liouville derivatives of the order $0 < \mathrm{Re}\,\alpha < 1, n = [\alpha]+1$ are given for a suitable function $f(x)$ by [**8**]:

(1) $\quad \partial_{a+}^{\alpha} f(t) = \frac{1}{\Gamma(n-\alpha)} \left(\frac{d}{dt}\right)^n \int_a^x \frac{f(\tau)d\tau}{(t-\tau)^{\alpha-n+1}}, \quad \partial_{b-}^{\alpha} f(t) = \frac{(-1)^n}{\Gamma(n-\alpha)} \left(\frac{d}{dt}\right)^n \int_x^b \frac{f(\tau)d\tau}{(\tau-t)^{\alpha-n+1}}.$

For the tensor analysis in oblique rectilinear coordinate system we extend the index notation and summation convention, implementing the case of fractional derivatives.

The order of derivative can assume only integer or half-integer values $\alpha = 1/2$. Fractional derivative, being applied twice, results in the common integer derivative.

Let us take up *partial* fractional derivatives in Minkowski space. The Greek indices appearing in this work, take on the four values $\sigma, \nu, \mu, \ldots = 0,1,2,3$. The Latin indices indicate spacelike components only and hence run $i, j, k, \ldots = 1,2,3$. Starting with the general definitions (1)-(2), we introduce the simplified notation for fractional partial Riemann-Liouville derivatives of the order $\alpha = 1/2$ with respect to coordinates $x_\sigma$:

| Fractional derivatives | left-hand | right-hand |
|---|---|---|
| semi-contravariant derivative | $f^{+,\sigma/2} \equiv [\partial_{-\infty,+} f(t)]^{\sigma/2}$ | $f^{-,\sigma/2} \equiv [\partial_{\infty,-} f(t)]^{\sigma/2}$ |
| semi-covariant derivative | $f^{+}{}_{,\sigma/2} \equiv [\partial_{-\infty,+} f(t)]_{\sigma/2}$ | $f^{-}{}_{,\sigma/2} \equiv [\partial_{\infty,-} f(t)]_{\sigma/2}$ |

**Table 1 Covariant and contravariant fractional derivatives**

We shall have the tensor field quantities located at a point, with various components referring to the axes at that point. When we change our coordinate system, the components will change according to the common laws, depending on the change of axes at the point considered. Let us see the effect of a change in the coordinate system on fractional derivatives. Take the new coordinates $x^{\tilde{\mu}}$, each a function of four $x^\sigma$. Making a small variation in $x^\sigma$, we get the four quantities forming the components of a contravariant vector. Being referred to the new axes, this vector has the components $\delta x^{\tilde{\mu}} = m_\nu^{\tilde{\mu}} \delta x^\nu$, where **m** is the transformation matrix to the new coordinates with components $m_\nu^{\tilde{\mu}} = \partial x^{\tilde{\mu}} / \partial x^\nu = x_{,\nu}^{\tilde{\mu}}$. The transformation rules for the fractional derivatives are

(2) $\qquad \tilde{f}^{\pm, \sigma/2} = \mu_\nu^{\tilde{\sigma}} f^{\pm, \nu/2}, \qquad\qquad \tilde{f}^{\pm}{}_{,\sigma/2} = \mu_{\tilde{\sigma}}^\nu f^{\pm}{}_{,\nu/2}.$

The transformation matrix $\boldsymbol{\mu}$ is positive square root of matrix $\mathbf{m}$ ([**9**], Ch. V), i.e. $m_\kappa^\sigma = \mu_\nu^\sigma \mu_\kappa^\nu$. The fractional derivatives are not tensors. We shall refer to the objects, which obey the fractional transformation rule (2), as *semi-tensors*. The semi-contravariant derivatives are semi-tensors of type $(1/2, 0)$, while the semi-covariant derivatives– of type $(0, 1/2)$. The definite combinations of semi-tensors do exhibit true tensor character.

At first, the fractional index of order $1/2$ may appear exactly twice either upstairs or downstairs. The upper or lower index, being applied twice, end up as partial derivatives

(3) $\qquad \left(f^{,\sigma/2}\right)^{,\sigma/2} \equiv \partial_{-\infty,+}^{\sigma/2}\left[\partial_{-\infty,+}^{\sigma/2} f(t)\right] = f^{,\sigma}, \qquad \left(f_{,\sigma/2}\right)_{,\sigma/2} \equiv \partial_{\sigma/2}^{\infty,-}\left[\partial_{\sigma/2}^{\infty,-} f(t)\right] = f_{,\sigma}.$

The quantities (3) transform as contravariant and covariant vectors respectively. Otherwise saying, quantities (8) are tensors of types $(1, 0)$ and $(0, 1)$.

The contraction of two fractional indices into an integer one is known as additive index law, or semi-group property of fractional operators. The resulting integer index (upstairs or downstairs) could be used for common summation with the corresponding opposite index (downstairs or upstairs). The fractional summation convention is applied as follows. The set of downstairs indices (integer or fractional) may be equal to an upstairs one. Then we have to sum over all values of this index of the same kind (integer or fractional). The index becomes a dummy. Then we have a tensor having two effective indices of the same kind less than the original one. Only quantities with balanced half-integer indices are of tensor character.

Secondly, twice-applied fractional index (one fractional upstairs index, one fractional downstairs index) leads to the scalar operators

(4) $\qquad \left(f^{-,\sigma/2}\right)_{,\sigma/2} \equiv \partial_{\sigma/2}^{\infty,-}\left[\partial_{\infty,-}^{\sigma/2} f\right], \qquad \left(f^{+,\sigma/2}\right)_{,\sigma/2} \equiv \partial_{\sigma/2}^{-\infty,+}\left[\partial_{-\infty,+}^{\sigma/2} f\right].$

The operators (4) have sense of "square roots" of Laplace operator. The operators are referred to as Riesz potential operator or operators of Riesz integral-differentiation. The operators possess remarkable properties, like isotropy and existence of variation principles.

The covariant derivatives of the contravariant components of the metric tensor vanish $g_{\lambda\kappa;\mu/2} = 0$, $g^{\lambda\kappa;\mu/2} = 0$. By the definition of connection the partial fractional derivative of the metric tensor is a linear and homogeneous form of the components of metric tensor [**10**]

$$g^{\lambda\kappa,\mu/2} = g^{\rho\kappa}\tilde{\tilde{\Gamma}}_\rho^{\mu\lambda} + g^{\lambda\sigma}\tilde{\tilde{\Gamma}}_\sigma^{\mu\kappa}, \qquad\qquad g_{\lambda\kappa,\mu/2} = g_{\rho\kappa}\tilde{\Gamma}_{\mu\lambda}^\rho + g_{\lambda\sigma}\tilde{\Gamma}_{\mu\kappa}^\sigma.$$

The solution of these linear equations delivers the objects

$$\tilde{\Gamma}_{\iota\lambda}^\mu = \frac{1}{2} g^{\mu\sigma}\left(g_{\sigma\iota,\lambda/2} + g_{\sigma\lambda,\iota/2} - g_{\iota\lambda,\sigma/2}\right), \qquad \tilde{\tilde{\Gamma}}_\mu^{\iota\lambda} = \frac{1}{2} g_{\mu\sigma}\left(g^{\sigma\iota,\lambda/2} + g^{\sigma\lambda,\iota/2} - g^{\iota\lambda,\sigma/2}\right),$$

which appear similar to Christoffel symbols

$$\Gamma_{\kappa\lambda}^\iota = \frac{1}{2} g^{\iota\mu}\left(g_{\mu\kappa,\lambda} + g_{\mu\lambda,\kappa} - g_{\kappa\lambda,\mu}\right), \qquad \Gamma_\mu^{\iota\lambda} = \frac{1}{2} g_{\mu\sigma}\left(g^{\sigma\iota,\lambda} + g^{\sigma\lambda,\iota} - g^{\iota\lambda,\sigma}\right)$$

The "fractional" Christoffel symbols contain the fractional derivatives of metric of order $1/2$. These symbols are neither tensors nor semi-tensors. The common link between two kinds of Christoffel symbols $\Gamma^{\lambda\iota}_{\kappa} = g^{\iota\sigma}\Gamma^{\lambda}_{\sigma\kappa}$ due to lack of Leibnitz property for fractional derivatives [**11**] is no more valid, such that $\widetilde{\Gamma}^{\lambda\iota}_{\kappa} \neq g^{\iota\sigma}\widetilde{\Gamma}^{\lambda}_{\sigma\kappa}$.

## 2.2 Representation of fractional derivatives by means of Fourier operators

The pseudodifferential operator might be represented in the form of integral Fourier operator [**12**]

(5) $\quad (\mathrm{D}f)(x) = F_{k \to x} d(k) F_{x \to k} f(x)$ .

Scalar function $d(k)$ is a symbol of operator $\mathrm{D}$, i.e. $d(k) = \mathrm{symbol}(\mathrm{D})$.

We use the following notation for direct and inverse Fourier transform $\hat{f}(k) = [F_{x \to k} f(x)](k)$, $f(x) = [F_{k \to x} \hat{f}(k)](x)$ for appropriate functions $f(x), \hat{f}(k)$ defined in $k \in R^n, x \in R^n$, $n = 4$.

The fractional derivatives may be considered as partial fractional derivatives with respect to space coordinates $x_\nu$ and momentum coordinates $k^\nu$. The Fourier transform of fractional derivative is the product of the Fourier transform of the operator and the Fourier transform of the function, such that

(6) $\quad \partial^{\alpha}_{-\infty,+} f(x) = [F_{k \to x}]\{(-ik)^{\alpha}[F_{x \to k} f(x)](k)\}(x)$.

The symbol of fractional derivative is $\quad \mathrm{symbol}(\partial^{\alpha}_{-\infty,+}) = (-ik)^{\alpha}$.

The symbol for "square root" of d'Alembert operator is

(7) $\quad \mathrm{symbol}(\Delta^{1/2}) = \sqrt{(-ik_\nu)(-ik^\nu)} \equiv i\sqrt{-k_0^2 + |k|^2}\qquad$ with $|k| = \sqrt{k_i k^i}$ .

## 2.3 Variational principles with fractional partial derivatives

*Fractional-Quadratic functionals*

Field equations with fractional derivatives could be deducted from variation principles. The Lagrangian of particular interest for our theory are those of the quadratic form. The starting point for derivation of variation principles is the Parseval equality:

(8) $\quad \int\limits_a^b f(t)(\partial^{\alpha}_{a+} g)(t) dt = \int\limits_a^b g(t)(\partial^{\alpha}_{b-} f)(t) dt.$

The Parseval equality (8) demonstrates, that $\partial^{\alpha}_{a+}$ is the adjoint operator to $\partial^{\alpha}_{b-}$. By choosing the function $g(t) = (\partial^{\alpha}_{b-} f)(t)$ and substituting it into Parseval equality (8), we define the fractional quadratic functional as

(9) $\quad S[f(t)] = \frac{1}{2}\int\limits_a^b \mathrm{L}(\partial^{\alpha}_{b-} f, \partial^{\alpha}_{a+} f) dt,$ where $\mathrm{L} = (\partial^{\alpha}_{b-} f)(\partial^{\alpha}_{a+} f)$.

The variation of fractional-quadratic functional (9) is

(10) $\quad \delta S[f(t)] = \int\limits_a^b (\partial^{2\alpha}_{a+} f + \partial^{2\alpha}_{b-} f) \delta f \, dt$ .

The fractional-quadratic functional for the function of several variables is written as

(11) $\quad S[f(x^\nu)] = \frac{1}{2}\int\limits_{R^n} (f^{+,\sigma/2})(f^{-},_{\sigma/2}) d\Omega.$

The first variation of $S[f(x^\nu)]$ is the bilinear form

$$(12) \quad \delta S[f, \delta f] = \frac{1}{2} \int_{R^n} \left\{ (f^-_{,\alpha/2})^{\alpha/2} + (f^+_{,\alpha/2})^{\alpha/2} \right\} \delta f \, d\Omega.$$

The admitted variations die out at infinity: $\delta f(\infty) = 0$. The functional (11) serves as variational functional for "square root" of Laplace operator.

## 2.4 Convolution and Fourier variational formulations

The investigation of symmetry properties of pseudodifferential equations is studied in most straightforward method, using the variational formulations. Consider an important case of spatially homogeneous symbol $d(k)$. The Green function of the operator D reads

$$d^*(x) = \left(\frac{1}{-2\pi i}\right)^{n/2} \int_{R^n} d(k) e^{ix_\alpha k^\alpha} d^n k.$$

There exist two equivalent variational formulations for a fractional pseudodifferential equation. The variational functional could be obtained by multiplying the fractional pseudodifferential equation $Df(x) = 0$ by $f(x)$ and integrating over $R^n$. The solution $f(x)$ of the fractional pseudodifferential equation is the unique minimizer of two form of variational functionals:

$$(13) \quad S_k[\hat{f}(k)] \to \min_{\hat{f}(k)}, \qquad S_x[f(x)] \to \min_{f(x)}$$

where

$$(14) \quad S_x[f(x)] = \left(\frac{1}{-2\pi i}\right)^{n/2} \int_{R^n \times R^n} d^*(x-y) f(x) f(y) d^n x d^n y, \quad S_k[\hat{f}(k)] = \int_{R^n} d(k) \hat{f}(k) \bar{\hat{f}}(k) d^n k.$$

The functional $S_x[f(x)]$ we call convolution functional, the $S_k[\hat{f}(k)]$ - variational functional Fourier. The equivalence of variational functionals is stated immediately

$$\left(\frac{1}{-2\pi i}\right)^{n/2} \int_{R^n \times R^n} d^*(x-y) f(x) f(y) d^n x d^n y = \int_{R^n} d(k) \hat{f}(k) \bar{\hat{f}}(k) d^n k.$$

Finally, the essential mathematical apparatus is assembled, and we proceed with formulation of physical hypothesis.

## 3 Governing equations of Riesz gravitation

### 3.1 Symmetries and conservation laws of fractional partial differential equations

The physically admissible symbols $d = d(\mathbf{k})$ obey several symmetry conditions:
(i) *Uniformity of space*. The symbol space must be invariant to shift of spatial coordinates. The symbol does not depend explicitly on spatial coordinates. This symmetry guarantees the momentum conservation.
(ii) *Uniformity of time*. The symbol does not depend explicitly upon time shift, what assurances energy conservation.
(iii) *Rotational symmetry of symbol* reassures angular momentum conservation.
(iv) *Isometric invariance*, or Lorenz invariance, essential for the relativistic conformity of the theory.

The symbol $d = d(\mathbf{k})$, depending only on momentum variables, satisfies the conditions (i)-(ii). The symbol $d = d\left(\sqrt{-k_0^2 + |k|^2}\right)$, depending only on the module of momentum variables, satisfies additionally the conditions (iii)-(iv). For brevity we put here, as usual, $c = 1$. If variational functional (16) satisfies some invariance condition, the same is valid for the corresponding symbol.

If the symbol is homogeneous function of $\sqrt{-k_0^2 + |k|^2}$, the governing equation is scale-invariant, like Newtonian gravity and electrodynamics. The generalized wave and static field equations possessing special invariant properties are summarized in Tables 1 and 2.

| Symbol | Symbolical Equation | Invariance |
|---|---|---|
| $l = -k_0^2 + |\mathbf{k}|^2 = \mathrm{symbol}(\square)$ | $\square \Phi(x) = \phi(x)$<br>D'Alembert Wave Equation | Spacial isotropy<br>*Scale invariant* |
| $\tilde{l} = -k_0^2 + |\mathbf{k}|^2 + \frac{1}{\lambda}\sqrt{-k_0^2 + |\mathbf{k}|^2} = \mathrm{symbol}(\tilde{\square})$ | $\tilde{\square}\Phi(x) = \phi(x)$<br>Riesz Wave Equation | Spacial isotropy |
| $\tilde{l}^{(\alpha)} = -k_0^2 + |\mathbf{k}|^2 + \frac{1}{\lambda^{1-\alpha/2}}\left(-k_0^2 + |\mathbf{k}|^2\right)^{\alpha/2} = \mathrm{symbol}(\tilde{\square}^{(\alpha)})$ | $\tilde{\square}^{(\alpha)}\Phi(x) = \phi(x)$<br>Generalized Riesz Wave Equation | Spacial isotropy |

**Table 2 Generalized wave equations and their symmetries**

| Symbol | Symbolical Equation | Invariance |
|---|---|---|
| $d = |\mathbf{k}|^2 = \mathrm{symbol}(\Delta)$ | $\Delta \Phi(x) = \varphi(x)$<br>Laplace Equation | Spacial isotropy<br>*Scale invariant* |
| $\tilde{d} = |\mathbf{k}|^2 + \frac{1}{\lambda}|\mathbf{k}| = \mathrm{symbol}(\tilde{\Delta})$ | $\tilde{\Delta}\Phi(x) = \varphi(x)$<br>Riesz Equation | Spacial isotropy |
| $\tilde{d}^{(\alpha)} = |\mathbf{k}|^2 + \frac{1}{\lambda^{2-\alpha}}|\mathbf{k}|^\alpha = \mathrm{symbol}(\tilde{\Delta}^{(\alpha)})$ | $\tilde{\Delta}^{(\alpha)}\Phi(x) = \varphi(x)$<br>Generalized Riesz Equation | Spacial isotropy |

**Table 3 Generalized potential equations and their symmetries**

### 3.2 Variational principles of Riesz gravitation

The role of gravitational potential in space-time coordinates plays metric $g_{\mu\nu}$. The source term in frame-independent formulation of the field equations is, as usual, the whole stress-energy tensor. Starting with the gravitational part, consider action for metric. The action of gravitational field contains the derivatives for metric of the order, not higher than one. Variational principle with this action leads to the equations of gravitational field, holding the derivatives for metric $g_{\mu\nu}$ of the order, not higher than two [**13**].

The extended action of gravitation is assumed to be the sum of Einstein-Hilbert action term and Riesz term:

(15) $S_g[g_{\mu\nu}] = S[g_{\mu\nu}] + \frac{1}{\lambda}\tilde{S}[g_{\mu\nu}],$

where $\lambda$ is a hypothetical metric parameter with the dimension of length. The Hilbert term reads

$$(16) \quad S[g_{\mu\nu}] = \int_\Omega \mathrm{L}\sqrt{-g}\,d\Omega, \qquad \mathrm{L} = g^{\iota\kappa}\left(\Gamma^\mu_{\iota\lambda}\Gamma^\lambda_{\kappa\mu} - \Gamma^\lambda_{\iota\kappa}\Gamma^\mu_{\lambda\mu}\right).$$

The Riesz term is

$$(17) \quad \tilde{S}[g_{\mu\nu}] = \int_\Omega \tilde{\mathrm{L}}\sqrt{-g}\,d\Omega, \qquad \tilde{\mathrm{L}} = g^{\iota\kappa}\left(\tilde{\Gamma}^\mu_{\iota\lambda}\tilde{\Gamma}^\lambda_{\kappa\mu} - \tilde{\Gamma}^\lambda_{\iota\kappa}\tilde{\Gamma}^\mu_{\lambda\mu}\right)$$

The scalar Lagrangian $\tilde{\mathrm{L}}$ is the fractional-quadratic function of the metric tensor components

$$4\tilde{L} = \left(g^{\mu\kappa,\alpha/2} + g^{\mu\alpha,\kappa/2} - g^{\kappa\alpha,\mu/2}\right)\left(g_{\alpha\kappa,\mu/2} + g_{\alpha\mu,\kappa/2} - g_{\kappa\mu,\alpha/2}\right) -$$
$$- \left(g^{\lambda,\iota/2}_\iota + g^{\lambda,\kappa/2}_\kappa - g^{\kappa,\lambda/2}_\kappa\right)\left(g^\nu_{\lambda,\nu/2} + g^\nu_{\nu,\lambda/2} - g^\mu_{\lambda,\mu/2}\right)$$

The variation of (15) with respect to metric $g_{\mu\nu}$ is performed accounting for Lorenz-Hilbert gauge conditions. To produce the correct right-hand side to the field equations in the presence of matter, the matter action is taken to be of the standard form

$$S_M[\phi; g] = \int_V \mathrm{L}_M \sqrt{-g}\,d^4x,$$

for Lagrangian density of matter $\mathrm{L}_M$. The stress-energy tensor is

$$T_{\alpha\beta} \equiv -2\frac{\partial \mathrm{L}_M}{\partial g_{\alpha\beta}} + \mathrm{L}_M g_{\alpha\beta}.$$

Then the field equations in the presence of matter follow from the variational principle

$$\delta\!\left(S[g_{\mu\nu}] + \lambda^{-1}\tilde{S}[g_{\mu\nu}] + S_M[\phi, g]\right) = 0.$$

### 3.3 Variational principles for weak Riesz gravitation

The coordinate frame of nearly Minkowskian coordinates the space-time metric is described by a small smooth perturbation $h_{\mu\nu}$ of nearly Minkowskian metric

$$\eta^{\mu\nu} = \eta_{\mu\nu} = \mathrm{diag}(-1,+1,+1,+1) = e_\mu \delta_{\mu\nu},$$

i.e. $g_{\mu\nu} = \eta_{\mu\nu} + h_{\mu\nu}$, $g^{\mu\nu} = \eta^{\mu\nu} - h^{\mu\nu}$, $\sqrt{-g} = 1 + h$, $h = e_\mu h_{\mu\mu}$, $h^{\mu\nu} = e_\mu e_\nu h_{\mu\nu}$.

We use Eisenhardt symbols $e_1 = e_2 = e_3 = 1$, $e_0 = -1$ for Minkowskian metric.

By the smallness we assume that in Taylor expansion of arbitrary function of the metric the higher order terms can be neglected. By smoothness we assume that also derivatives of the perturbation are irrelevant of the same order. For the same order of magnitude of the coordinate perturbation and the perturbations of the Minkowski metric action (15) reduces to

$$(18) \quad \sigma_g[h_{\mu\nu}] = \sigma[h_{\mu\nu}] + \frac{1}{\lambda}\tilde{\sigma}[h_{\mu\nu}]$$

Hilbert term of Lagrangian of the desired order of magnitude becomes

$$\sigma[h_{\mu\nu}] = \frac{1}{2}\int_\Omega \left[h^{\mu\nu}_{,\mu}h_{,\nu} - h^\mu_{\sigma,\rho}h^{\rho\sigma}_{,\mu} + \frac{1}{2}\eta^{\mu\nu}h_{\rho\sigma,\mu}h^{\rho\sigma}_{,\nu} - \frac{1}{2}\eta^{\mu\nu}h_{,\mu}h_{,\nu}\right]d\Omega.$$

Fractional-quadratic functional (18) reads

$$\tilde{\sigma}[h_{\mu\nu}] = \frac{1}{2}\int_\Omega \left[\left(h^\mu_\nu\right)^+_{,\mu/2}h^{-,\rho/2} - \left(h_{\nu\rho}\right)^+_{,\mu/2}\left(h^{\mu\nu}\right)^{-,\rho/2} + \frac{1}{2}\left(h_{\rho\sigma}\right)^+_{,\mu/2}\left(h^{\rho\sigma}\right)^{-,\mu/2} - \frac{1}{2}h^+_{,\mu/2}h^{-,\mu/2}\right]d\Omega.$$

### 3.4 Equation of weak gravitations field

Performing variation of $\sigma[h_{\mu\nu}]$ we obtain common linear wave equations. Hilbert gauge condition reduces equations to D'Alembert wave equations

(19) $\Box h_{\mu\nu} \equiv h_{\mu\nu,\sigma}{}^{,\sigma} \equiv F^{-1}_{\xi \to x} l(\xi) F_{x \to \xi} h_{\mu\nu} = 0$   with   $\text{symbol}(\Box) = l(k) \equiv -\frac{1}{c^2} k_0^2 + |k|^2$.

Presuming static Hilbert-Lorenz gauge conditions, the static Newtonian potentials in vacuum satisfy Laplace equation

(20) $\Delta h_{\mu\nu} \equiv h_{\mu\nu,i}{}^{,i} = F^{-1}_{k \to x} d(k) F_{x \to k} h_{\mu\nu} = 0$   with   $\text{symbol}(\Delta) = d(k) \equiv |k|^2$.

In the static case gravitational potential obeys the equation $g_{00} = 1 + h_{00} = 1 + 2c^{-2}\phi$.

Performing variation of functional (18) we obtain linear equations of Riesz gravitation. The equations for potentials in Riesz theory are partial pseudo-differential equations. The symbols of the equations read as

(21) $\text{symbol}(\tilde{\Box}) = \tilde{l}(k) \equiv -\frac{1}{c^2} k_0^2 + |k|^2 + \frac{1}{\lambda}\sqrt{-\frac{1}{c^2} k_0^2 + |k|^2}$,

(22) $\text{symbol}(\tilde{\Delta}) = \tilde{d}(k) \equiv |k|^2 + \frac{1}{\lambda}|k|$.

The linear static, spatially isotropic pseudo-differential equation

(23) $\tilde{\Delta} h_{\mu\nu}(x) = F^{-1}_{k \to x} \tilde{d}(k) F_{x \to k} h_{\mu\nu} = 0$

with the symbol (22) will be referred to as Riesz potential operator.
The isometric-invariant, spatially isotropic pseudo-differential equation

(24) $\tilde{\Box} h_{\mu\nu}(x) = F^{-1}_{k \to x} \tilde{l}(k) F_{x \to k} h_{\mu\nu} = 0$

with symbol (21) is the Riesz wave equation in vacuum.

## 4 Green functions for weak gravitational field

### 4.1 Green functions and asymptotics of weak Riesz gravitation field

The Green functions $\tilde{\Phi}(x)$ and $\tilde{\Psi}(x)$ of pseudo-differential operators $\tilde{\Box}$ и $\tilde{\Delta}$ are found with integral transformation method [**14**]. At first, we find the static fundamental solution for operator $\tilde{\Delta}$. A fundamental solution for the operator $\tilde{\Delta}$ is a distribution $\tilde{\Phi}(x)$, so that $\tilde{\Delta}\tilde{\Phi} = \delta(x)$. Its Fourier transformation satisfies $\tilde{d}(-ik) F_{x \to k} \tilde{\Phi} = 1$.

*The fundamental solution for negative metric parameter*

To determine the fundamental solution for $\lambda < 0$, we perform the inversion of Fourier transformation. Understanding the integral as its Cauchy principal value, we get the fundamental solution:

(25) $\tilde{\Phi}(r, \lambda < 0) = \frac{2\pi}{r}\left[\pi \cos\left(\frac{r}{|\lambda|}\right) + 2\text{Si}\left(\frac{r}{|\lambda|}\right)\cos\left(\frac{r}{|\lambda|}\right) - 2\text{Ci}\left(\frac{r}{|\lambda|}\right)\sin\left(\frac{r}{|\lambda|}\right)\right]$.

where $r = |x| = \sqrt{x_i x^i}$.

*The fundamental solution for positive metric parameter*

The Green function of Riesz potential for $\lambda > 0$ demonstrates quite different character. We get

$$(26) \quad \tilde{\Phi}(r, \lambda > 0) = \frac{2\pi}{r}\left[\pi \cos\left(\frac{r}{\lambda}\right) - 2\operatorname{Si}\left(\frac{r}{\lambda}\right)\cos\left(\frac{r}{\lambda}\right) + 2\operatorname{Ci}\left(\frac{r}{\lambda}\right)\sin\left(\frac{r}{\lambda}\right)\right].$$

Fundamental solution (26) turns into the Newtonian potential for $\lambda \to \infty$. At small distances $r \ll \lambda$ the deviation between the fundamental solution of for Newtonian potential and Riesz potential is of the order of $4\pi\lambda^{-1}(\gamma - 1 + \ln(r/\lambda))$.

## 4.2 Gravitational potential of a point mass

We calculate now, using (25), the gravitation potential and acceleration of a point mass $M$. The common dimensional factor $\kappa = -GM/2\pi^2$ is used with Newtonian constant of gravitation $G$. The normalized Newtonian potential reads $\kappa\Phi(r) = -GM/r$. The Riesz potential $\kappa\tilde{\Phi}(r, \lambda < 0)$ oscillates and at the extremum point $r_e = 0.313\lambda$ is 1.29.. of Newtonian potential.

The Riesz gravitational acceleration of the point mass $M$ is

$$\tilde{A}(r) = \frac{\kappa}{\lambda r}\left[\sin\left(\frac{r}{|\lambda|}\right) + \frac{2}{\pi}\operatorname{Ci}\left(\frac{r}{|\lambda|}\right)\cos\left(\frac{r}{|\lambda|}\right) + \frac{2}{\pi}\operatorname{Si}\left(\frac{r}{|\lambda|}\right)\sin\left(\frac{r}{|\lambda|}\right)\right] +$$

$$+ \frac{\kappa}{r^2}\left[\pi\cos\left(\frac{r}{|\lambda|}\right) + 2\operatorname{Si}\left(\frac{r}{|\lambda|}\right)\cos\left(\frac{r}{|\lambda|}\right) - 2\operatorname{Ci}\left(\frac{r}{|\lambda|}\right)\sin\left(\frac{r}{|\lambda|}\right)\right].$$

The asymptotic behavior of the Riesz acceleration reads

$$\tilde{A}(r) \underset{r/\lambda \to \infty}{=} \frac{2\kappa}{\lambda r}\sin\left(\frac{r}{|\lambda|}\right) + \frac{2\kappa}{r^2}\cos\left(\frac{r}{|\lambda|}\right) + O\left(\frac{1}{r^3}\right).$$

The Riesz acceleration achieves also its maximum value at the distance $r_m \approx 1.62\lambda$. On this distance, the value of Riesz gravitational acceleration is 2.64.. of Newtonian gravitational acceleration $A(r) = \kappa/r^2$, i.e. $\tilde{A}(r_m)/A(r_m) = 2.64...$

The ratios of Riesz potential to Newtonian potential $\tilde{\Phi}/\Phi$ and Riesz acceleration to Newtonian acceleration $\tilde{A}(r)/A(r)$ as function of radius $r$ are shown on **Fig.1** and **Fig. 2**. The accelerations' ratio is shown by solid line, and the potentials' ratio is plotted by dashed line.

At the small distances the potentials and accelerations matches, such that

$$\frac{\tilde{A}(r)}{A(r)} \underset{r/\lambda \to 0}{=} 1 + \frac{2}{\pi}\frac{r}{\lambda} + O(r^2),$$

$$\frac{\tilde{\Phi}(r)}{\Phi(r)} \underset{r/\lambda \to 0}{=} 1 - \frac{2}{\pi}\frac{r}{\lambda}\left(\gamma - 1 + \ln\frac{r}{\lambda}\right) + O(r^2).$$

Remarkably, that on the large distances the ratio of the accelerations could be arbitrary high, because

$$\frac{\tilde{A}(r)}{A(r)} \underset{r/\lambda \to \infty}{=} 2\frac{r}{\lambda}\sin\left(\frac{r}{\lambda}\right) + 2\cos\left(\frac{r}{\lambda}\right) + O\left(\frac{1}{r}\right).$$

The potentials' ratio remains limited, such that the amplitude of Riesz potential is equal the value of Newtonian potential at the observation point:

$$\frac{\tilde{\Phi}(r)}{\Phi(r)} \underset{r/\lambda \to \infty}{=} 2\cos\left(\frac{r}{\lambda}\right) - \frac{2}{\pi}\frac{\lambda}{r} + O\left(\frac{1}{r^3}\right).$$

The radius of the first attraction zone is $R_s \approx 2.6\lambda$. With the rough guess of metric parameter $\lambda \approx 7.5\,kpc = 0.231 \cdot 10^{21} m$, the radius of first attraction zone is $R_s \approx 19.5\,kpc \approx 0.6 \cdot 10^{21} m$. This corresponds to the maximal size of observable galaxies of $40-50\,kpc$. The objects with the diameters more than $2R_s$ in the Riesz gravitation will be unstable. There exist also the outer attractions zones $6 < r/\lambda < 9$, $13 < r/\lambda < 15$ etc (see **Fig.2**). In the following chapter we try to interpret the outer attraction zones as the belts of neighboring dwarf galaxies.

## 5  On the possible experimental tests of post-Newtonian gravity

### 5.1  The Solar system orbital effects induced by Riesz acceleration

The direct verification of the weak-field Newtonian limit substantially based on Solar system experiments. In these experiments, the masses of gravitating bodies could be good estimated and the distances and velocities of bodies being continuously measured with high precision. The experimental data was collected and revised for centuries. Solar system experiments that test the weak-field, post-Newtonian limit of metric theories strongly favor general relativity [**15**]. We will test now, weather the Riesz theory violates the contemporary experimental data for Solar system dynamics and what new, possibly observable, secular effects it predicts.

The powerful instrument for verification of gravitation laws delivers the perturbation theory. On the sizes of Solar system, the difference between the Riesz acceleration and the Newtonian one is only small quantity. We evaluate now the orbital perturbations induced by a disturbing acceleration of Riesz term on the Keplerian orbital elements of Solar system members. Well known, that the semimajor axis, the eccentricity, the perihelion and the mean anomaly of plants are affected by short-periods signatures. The perihelion and the mean anomaly undergo also long-period, secular effects. This secular effect could be acquired with high precision by contemporary astronomical methods [**16**].

The Gauss equations for the variations of the semimajor axis $a$, the eccentricity $e$, the inclination $i$, the longitude of the ascending node $\Omega$, the argument of pericentre $\omega$ and the mean anomaly $\mathrm{M}$ of a test particle of mass $m$ in the gravitational field of a body $M$ can be reduced from

(27) $\ddot{\mathbf{r}} + \dfrac{\mu}{r^3}\mathbf{r} = \mathbf{A}$,

where $\mu = G(m+M)$, $\mathbf{r} = \mathbf{r}_m - \mathbf{r}_M$ ([**17**]).

The equation (27) holds for every perturbing acceleration $\mathbf{A}$, whatever its cause or size. The variations of the elements are, for an entirely radial acceleration $A_r$

(28) $\dfrac{da}{dt} = \dfrac{2e}{n\sqrt{1-e^2}} A_r \sin f$,    $\dfrac{de}{dt} = \dfrac{\sqrt{1-e^2}}{na} A_r \sin f$,    $\dfrac{d\omega}{dt} = -\dfrac{\sqrt{1-e^2}}{nae} A_r \cos f$,

(29) $\dfrac{d\mathrm{M}}{dt} = n - \dfrac{2}{na} A_r - \sqrt{1-e^2}\dfrac{d\omega}{dt}$,    $\dfrac{di}{dt} = 0$, $\dfrac{d\Omega}{dt} = 0$.

In the expressions (28)-(29) the following definitions are used: $n = 2\pi/P$ is the mean motion, $P$ is the test particle's orbital period, $f$ is the true anomaly counted from the pericentre, $p = (1-e^2)a$ is the semilactus rectum of the Keplerian ellipse and $A_r$ is the projection of the perturbing acceleration **A** on the radial direction of the co-moving frame $\{\mathbf{r},\mathbf{t},\mathbf{n}\}$. The inclination $i$ and the node $\Omega$ are not perturbed by entirely radial perturbing acceleration.

|  | Semimajor axis $a$, $m \cdot 10^{-11}$ | Period, $s \cdot 10^{-9}$ | Eccentricity | Newtonian Acceleration of Sun $A(r)$, $m\,s^{-2}$ | Difference to Newtonian acceleration $\tilde{A}(r)/A(r)-1$ |
| --- | --- | --- | --- | --- | --- |
| Mercury | 0.57909 | 0.0076005 | 0.2056 | 0.0397 | $0.16 \cdot 10^{-9}$ |
| Venus | 1.08209 | 0.019414 | 0.0068 | 0.01139 | $0.29 \cdot 10^{-9}$ |
| Earth | 1.49598 | 0.031558 | 0.0167 | $5.96 \cdot 10^{-3}$ | $0.40 \cdot 10^{-9}$ |
| Mars | 2.2794 | 0.059355 | 0.0934 | $2.56 \cdot 10^{-3}$ | $0.62 \cdot 10^{-9}$ |
| Jupiter | 7.7832 | 0.374335 | 0.0485 | $2.2 \cdot 10^{-4}$ | $0.214 \cdot 10^{-8}$ |
| Saturn | 14.27 | 0.92959 | 0.0556 | $6.55 \cdot 10^{-5}$ | $0.392 \cdot 10^{-8}$ |
| Uranus | 28.696 | 2.6512 | 0.0472 | $1.6200 \cdot 10^{-5}$ | $0.789 \cdot 10^{-8}$ |
| Neptune | 44.96 | 5.200 | 0.0086 | $0.6597 \cdot 10^{-5}$ | $0.123 \cdot 10^{-7}$ |
| Pluto | 59.0 | 7.816 | 0.25 | $0.3832 \cdot 10^{-5}$ | $0.162 \cdot 10^{-7}$ |

**Table 4 Elements of planet orbits and Sun accelerations**

The difference of Riesz acceleration to the Newtonian monopole term ranges on the distances of Solar system from so that the equations can be treated in the perturbative way.

The right hand-sides of (28)-(29) have to be evaluated on the unperturbed Keplerian ellipse. The secular effects can be obtained by averaging over one orbital period the right hand-sides of the (28)-(29). Instead of the true anomaly $f$ we use the eccentric anomaly $E$, defined as $M = E - e\sin E$. The equations for secular variables in terms of E are

$$(30) \quad \frac{dE}{dt} = \frac{n}{1-e\cos E}, \qquad \frac{r}{a} = 1 - e\cos E, \quad \cos f = \frac{\cos E - e}{1-e\cos E}, \quad \sin f = \frac{\sqrt{1-e^2}\sin E}{1-e\cos E}.$$

The shifts over one full orbital revolution, i.e. from $E_0$ to $E_0 + 2\pi$, vanish for the semimajor axis $a$ and the eccentricity $e$. Consequently, for such Keplerian orbital elements there are no secular orbital effects.

The pericentre and the mean anomaly demonstrate secular orbital effects. Secular rates of the pericentre and the mean anomaly are

$$(31) \quad \left\langle \frac{d\omega}{dt} \right\rangle = \frac{\sqrt{1-e^2}}{a\,n} A_r, \qquad \left\langle \frac{dM}{dt} \right\rangle = -\frac{3}{a\,n} A_r.$$

From Table 1 is apparent, that for Solar system the Riesz potential (26) deviates from Newtonian only slightly. The difference between Riesz potential and Newtonian potential could be interpreted as perturbation $\delta\Phi(r) = \tilde{\Phi}(r) - \Phi(r)$. The elegant expression for secular rate of perihelion precession could be obtained for nearly circular orbits $e \ll 1$. The secular anomalous (non-Einsteinian) perihelion precession for one orbiting period $P$ is given by

$$(32) \quad \Delta\omega = \pi u \frac{\partial^2 \delta\Phi(1/u)}{\partial u^2} \left[ \frac{\partial \delta\Phi(1/u)}{\partial u} \right]^{-1} \bigg|_{u=1/a},$$

for the given semimajor axis of orbit $a$ and $u = 1/r$.

The asymptotic expansion of (32) delivers the simple correction formula for secular anomalous (non- Einsteinian) perihelion precession predicted for one orbiting period

$$(33) \quad \Delta\omega = -\frac{\pi^2}{4}\frac{a}{\lambda}.$$

The expression (32) delivers for nearly circular orbits the value of centennial General Relativity precession rate $\Delta\omega_{GR} = 3\pi r_g / a$, were $r_g = 2GM/c^2$ gravitational radius of the Sun and $M = 2 \cdot 10^{30} kg$ mass of Sun. More accurate estimation [18] of the precession rate for one orbiting period is

$$(34) \quad \Delta\omega_{GR} = \frac{3\pi r_g}{a(1-e^2)}.$$

The perihelion precession rate (34) decreases with the growing radius of orbit. Otherwise, the anomalous perihelion precision rate due to Riesz correction (33) increases with the orbit radius $a$. Centennial General Relativity precession rate (34) for Mercury is thousand times centennial anomalous precession rate (33). For the outer planets (Uranus, Neptune and Pluto) dominates the anomalous precession due to Riesz correction (see Table 5). According to the latest published results [19], this effect is yet too tiny to being detected with contemporary methods.

**Table 5 Secular anomalous precision, perihelion rate and mean anomaly of planets**

|  | Centennial anomalous precession rate (33), arcsec | Centennial General Relativity precession rate (34), arcsec | Centennial precession rate due to Sun quadrupole effect [20], arcsec | Centennial precession rate due to gravitomagnetic Lense–Thirring effect [21], arcsec |
|---|---|---|---|---|
| Mercury | -0.053 | -43.11 | $4.7 \cdot 10^{-2}$ | $-2 \cdot 10^{-3}$ |
| Venus | -0.0386 | -8.65 | $5 \cdot 10^{-3}$ | $-3 \cdot 10^{-4}$ |
| Earth | -0.0328 | -3.85 | $1.6 \cdot 10^{-3}$ | $-1 \cdot 10^{-4}$ |
| Mars | -0.0266 | -1.35 | $3 \cdot 10^{-4}$ | $-3 \cdot 10^{-5}$ |
| Jupiter | -0.0144 | -0.0625 | $5 \cdot 10^{-6}$ | $-7 \cdot 10^{-7}$ |
| Saturn | -0.01064 | -0.01374 | $6 \cdot 10^{-7}$ | $-1 \cdot 10^{-7}$ |
| Uranus | -0.0075 | -0.002394 | $5 \cdot 10^{-8}$ | $-1 \cdot 10^{-8}$ |
| Neptune | -0.0059 | -0.000777 | $1 \cdot 10^{-8}$ | $-5 \cdot 10^{-9}$ |

**Table 6 Comparison of centennial precision rates**

## 5.2 The Riesz effects in the Galaxy dynamics

Consider the rotation curve of the Galaxy (**Fig. 3**, [22]). The linear speed of the outer objects in halo of Galaxy demonstrates, that in the distant regions beside the observable gas exists the considerable amount of dark matter. The dark matter, has to be non-baryonic and that its effects are observed in stellar systems. Being completely decoupled dark and normal matter can mix in any ratio to form the objects in the observable universe, and indeed astronomical data show the relative content of dark matter to vary dramatically from object to object.

In an attempt to avoid the need for dark matter several modifications of the law of gravity have been proposed in the past decades. The modifications of Newtonian dynamics as an alternative to non-baryonic dark matter were proposed after realizing that mass discrepancies are observed in stellar systems. The fundamental theoretical problem for some modifications of Newtonian dynamics is the violation of conservation laws and essentially non-relativistic, non-covariant character of the equations.

Remarkably, that "dark matter" hypothesis is not required for the Riesz gravitation potential. The explanation of this phenomenon lay in nature of Riesz potentials. We start with the estimation of metric constant, attempting to match it to observable data. The analysis of star population with definite density maximum at this radius occasioned much thought. The metric constant merely corresponds to the size of galaxies. The roughly estimated value of metric constant is $\lambda \approx 7.5\, kpc = 0.23 \cdot 10^{21}\, m$. Not accidentally, the distance from Sun to the center of Galaxy is also nearly $\lambda$. On the distance of $r_e \approx 4 \cdot 10^{20}\, m$ the acceleration of Riesz gravitational field, as shown in previous section, is 2.6.. higher that the acceleration of Newtonian gravitational field from the same mass. The Riesz potential demonstrates the concentric regions of stability and instability. In the Riesz model, outer regions of galaxies with radiuses more than $3\lambda$ are gravitationally unstable. The medium in the outer regions of galaxies will be lost and forms the dwarf satellite galaxies in the belts of stabilities. This explains also a notable fact, that the maximal radius of observable galaxies is also about the order of 20 kpc. The neighboring galaxies of Local Group are most probably orbiting in the regions of stability (**Fig. 4**). The following distances of the nearest Local Group Galaxies from the center of Milky Way were used [**23,24**]: $r(CMAdw) = 42\, kly$, $r(SagDEG) = 68\, kly$, $r(LMC) = 179\, kly$, $r(SMC) = 210\, kly$.

## 6 Deflection of light and gravitational lensing

### 6.1 Deflection of light

Gravitation lensing delivers an additional instrument for the characterization and testing of gravitation potential on the galactic scales [**25**]. All the matter by between the light source and the observer affects the path, the size and the cross section of a propagating light bundle. For most practical purposes, we can assume that the lensing action is dominated by a single matter inhomogeneity M at some location between source S and observer O (**Fig.5**). This is usually called the 'thin lens approximation': All the action of deflection is thought to take place at a single distance. This approach is valid only if the relative velocities of lens, source and observer are small compared to the velocity of light $v \ll c$ and if the gravitation potential is small. These two assumptions are justified in all astronomical cases of interest. This ``lens thickness" is small compared to the typical distances between observer and lens or lens and background quasar/galaxy, respectively. In view of the simplifications just discussed, we can describe light propagation close to gravitational lenses in a locally Minkowskian spacetime perturbed by the gravitational potential of the lens to first post-Newtonian order. We now evaluate the deflection angle of a point mass $M$. The effect of spacetime curvature on the light paths can then be expressed in terms of an effective index of refraction $n_G$, which is given by $n_G = g_{00} = 1 + 2\kappa c^{-2} \tilde{\Phi}$. Note that the potential is negative if it is defined such that it approaches zero at infinity. As in normal geometrical optics, a refractive index $n > 1$ implies that light travels slower than in free vacuum. Thus, the effective speed of a ray of light in a gravitational field is $c_{eff} = c / n_G$.

For thin lens approximation, the deflection of light ray is the integral along the light path of the gradient of $n_G$ perpendicular to the light path. Observer and source assumed to be infinity. The deflection of light $\tilde{\alpha}, \alpha$ for Riesz and Newtonian potentials correspondingly read

$$(35) \quad \tilde{\alpha} = \frac{2}{c^2} \int_{Source}^{Observer} \frac{b\nabla\tilde{\Phi}}{\sqrt{b^2+z^2}} dz = \frac{r_g}{\lambda}\tilde{\vartheta}(\theta), \quad \alpha = \frac{2}{c^2}\int_{Source}^{Observer}\frac{b\nabla\Phi}{\sqrt{b^2+z^2}}dz = \frac{r_g}{\lambda}\vartheta(\theta),$$

where $b$ is the impact parameter of the unperturbed light ray, and $z$ indicates distance along the unperturbed light ray from the point of closest approach, $\theta = b/\lambda$. To deal with dimensionless deflection functions, we use the metric parameter $\lambda$ as scale coefficient. The dependence of light deflection as function of impact parameter from a source is given by dimensionless deflection functions $\tilde{\vartheta}(\theta), \vartheta(\theta)$. The deflection function of a point or spherical mass for Newtonian potential is $\vartheta(\theta) = 1/\theta$ (See, for instance, [**26**]).

For Riesz potential the deflection as a function of parameter $\theta$ could be obtained numerically and is shown on **Fig.6**. If the impact parameter is much smaller than the metric parameter, $b << \lambda$, we have $\tilde{\vartheta}(\theta) \underset{\theta\to 0}{\to} \vartheta(\theta)$. For small distances (as $\theta \to 0$) the asymptotic estimate of both deflection functions are the same. Thus, both potentials demonstrate the same lensing patterns on small distances, and the expected microlensing effects must be essentially the same for both potentials.

Another pattern occurs, when the impact parameter of the value or exceeds the metric parameter $\lambda$. The effects on the large scales, i.e. for parameter values $\theta > 1$, are radically different. For investigation of the asymptotic behavior of deflection function, we must work out the suitable analytical representation for this function. To ease the task, we utilize the asymptotic formula for the Riesz acceleration as a function of $r$ is

$$\nabla\tilde{\Phi} \cong 2GM(r\lambda)^{-1}\sin(r/\lambda) = 2c^2 r_g (r\lambda)^{-1}\sin(r/\lambda) \text{ for } r > \lambda.$$

An integral representation for the deflection function (35) reduces to

$$(36) \quad \tilde{\vartheta}(\theta) \cong \tilde{\vartheta}_*(\theta) = \int_1^\infty \frac{2\sin(\theta u)}{u\sqrt{u^2-1}}du.$$

Properly speaking, we need only the estimation for asymptotic representation of the function (36) for large values of $\theta$. The asymptotic behavior of (36) will be obtained using the methods for asymptotic expansion of integrals with oscillation kernels [**27**]. The integral (35) is a product of oscillation function $\sin(\theta u)$ and the kernel $k(u) = 2u^{-1}(u^2-1)^{-1/2}$. The asymptotic essential component of the kernel in the singularity point $u=1$ determines the asymptotic expansion of the integral for large values of $\theta$. The power series expansion of the kernel reads $k(u) = k_0(u) + O((u-1)^{1/2})$, $k_0(u) = \sqrt{2}(u-1)^{-1/2}$ at the point $u=1$.

For large values of $\theta$ the deflection function possess the asymptotic behavior (**Fig.7**)

$$(37) \quad \tilde{\vartheta}(\theta) \approx \tilde{\vartheta}_*(\theta) \underset{\theta\to\infty}{\sim} \int_1^\infty k_0(u)\sin(\theta u)du = \sqrt{\frac{2\pi}{\theta}}\cos(\theta - \pi/4).$$

Deserves mention, that the integral (40) could be also expressed in terms of generalized hypergeometric functions ${}_\alpha F_{\beta,\gamma}$ [**28**]. The deflection function (37) of the point mass $\tilde{\vartheta}(\theta)$ oscillates. In the central zone $b < 2\lambda$, where the deflection function is positive, the light deflects towards the central mass. In the regions, where the deflection function is negative, the light beam deflects outwards the mass. The predicted deflection of light ray is stronger, that the Newtonian. At $b = \lambda$ we have $\tilde{\vartheta}(b) \approx 2\vartheta(b)$. The Newtonian deflection function declines on the infinity as $\theta^{-1}$, while the Riesz deflection function vanishes slowly, as $\theta^{-1/2}$. Thus, the deflection of light beam caused by the same heavy mass on large distances in Riesz potential theory is much higher, the deflection of light ray in Newtonian potential theory. This behavior explains the observable high values of light deflection due to the galaxy clusters and the necessity of higher amounts of dark matter for the classical explanation of this light deflection. Some known peculiarities of lensing could be explained by the oscillation and slow decay of deflection function. Well known, that gravitational lensing by clusters of galaxies causes images of distant galaxies to be distorted and often split into multiple images. The gravitational mass of the lensing cluster, and its distribution, can be recovered through detailed analysis of the image pattern surrounding the cluster. Because of slow decay of deflection function, the lenses show a far more extended spatial extent than the visible cluster.

## 6.2 Equations of gravitational lensing and multiplicity of images

The equation of lens is similar to the classical one. Consider the thick curve on the Fig. 5, which connects the source S and the lower image $O_1$. The point heavy mass M deflects the light. The impact parameter is the distance from the mass M to the straight line $SO_1$. Denote the distance from the mass M to the straight line $SO$ as $r_0$. Comparison of similar triangles delivers

$$(38) \quad \frac{b - r_0}{D_{OO_1}} = \frac{D_{MS}}{D_{OS}}.$$

The distances between points $S, O, O_1, O_2, M$ are denoted as $D_{OO_1}, D_{OO_2}, D_{OS}, D_{OM}, D_{MS}$. Remembering, that $D_{OO_1} = \tilde{\vartheta}(b/\lambda) D_{OM} r_g / \lambda$ and $D_{OS} = D_{OM} + D_{MS}$, we get from (38) the common lens equation for gravitational lens

$$(39) \quad \frac{r_g}{\lambda} \tilde{\vartheta}\left(\frac{b}{\lambda}\right) \frac{1}{b - r_0} = \frac{1}{D_{MS}} + \frac{1}{D_{OM}}.$$

Consider the case of infinitely remote source. In this case $r_0 \to 0$ and $D_{MS} \to \infty$. In the classical gravity, the beam light from the infinitely remote source focuses at the distance from the heavy mass M, such that the single focus length of the gravitational lens is $R_f \equiv D_{OM} = b^2 r_g^{-1}$. In the Riesz gravity, there exist several solutions for the lens equations. The focus lengths are given by the relation

$$(40) \quad \tilde{R}_g \cong b^{3/2} \lambda^{1/2} \left[ r_g \cos\left(\frac{b}{\lambda} - \frac{\pi}{4}\right) \right]^{-1}.$$

The oscillating behavior of light deflection causes periodical ring patterns around the center of the image even for the point heavy mass.

## 7 The propagation and diffusion of gravitational waves

### 7.1 Fundamental solution of Riesz wave operator

Similar technique is applied to Riesz wave operator $\tilde{\Box} h_{00}(x,t) = \delta(x,t)$. The fundamental solution of reads

$$(41) \quad \tilde{\Psi}(x,t) = F^{-1}_{k \to x}[L^{-1}_{p \to t} \tilde{\Psi}(p,k)], \quad \tilde{\Psi}(p,k) = \left( p^2 + |k|^2 + \lambda^{-1}\sqrt{p^2 + |k|^2} \right)^{-1}$$

where $L^{-1}_{p \to t}[f(p)]$ is the inverse Laplace transform with respect to time.

We need the expression for fundamental solution in the far zone: $p^2 + |k|^2 \ll \lambda^{-2}$.

For this purpose we keep in image only the term with $\lambda^{-1}$. The image (41) reduces into $\tilde{\Psi}(p,k) = \left( \lambda^{-1}\sqrt{p^2 + |k|^2} \right)^{-1}$. Inverse Laplace transformation of the image delivers $L^{-1}_{p \to t}[\tilde{\Psi}(p,k)] = \lambda J_0(|k|t)$. The essential component of fundamental solution in the far zone $p^2 + |k|^2 \ll \lambda^{-2}$ reads

$$(42) \quad \tilde{\Psi}(x,t) \underset{r \to \infty}{\propto} F^{-1}_{k \to x}[\lambda J_0(|k|t)] = 2\lambda(t^2 - r^2)^{-\frac{3}{2}} \theta(t-r).$$

In classical General Relativity, gravitational waves obey the D'Alembert wave equation with the fundamental solution

$$(43) \quad \Psi(x,t) = \theta(t) F^{-1}_{k \to x}\left[ \frac{\sin(|k|t)}{|k|} \right] \equiv \frac{\theta(t)}{2\pi} \delta(t^2 - |x|^2).$$

Here $\theta(t)$ is Heaviside function.

### 7.2 Diffusion of wavefront for gravitational waves

There is a fundamental difference between the Green's function for Riesz wave equation and the Green's function for D'Alembert wave equation. Namely, the solutions of Riesz (42) and D'Alembert wave equations (43) exhibit two principally distinct characters of wave propagation. The Green's function of D'Alembert wave equation is a Dirac delta-function $\delta_{S_t}(x)$ modulated by the geometrical spreading $(4\pi t)^{-1}$. This means that the response to a $\delta(x)$-function source has the same shape as the input $\delta(t)$-function, that excites the wave field. An impulsive input leads to an impulsive output with a time delay given by $t = x$ and the solution is only nonzero at the wave front $K_t = \{|x| = t\}$. In other words, the solution for D'Alembert wave equation comply with the Huygens principle.

Contrarily, expression (42) shows that an impulsive input in Riesz wave equation in space leads to a response that is not impulsive. The response has an infinite duration and decays with time as $(t^2 - r^2)^{-\frac{3}{2}}$. The solution is not only nonzero at the wave front $K_t = \{|x| = t\}$, but it is nonzero everywhere within this wave front. This means that for Riesz wave equation an impulsive input leads to a sound response that is of infinite duration. One can therefore say that, in Riesz theory, any emitted gravitation wave will reverberate forever. The solution for Riesz wave equation does not obey the Huygens principle. The diffusion of Riesz wave leads to the blur of the peak on the wave front and decay of the wave amplitude on the leading edge of sharp wave front. The equivalent characteristic time of sharp wave front disruption is $\tau = \lambda/c$.

We try to explain now one known astrophysical phenomenon, namely the lack of *received* gravitation wave. As the size of binary pulsar is of several AE, the emission of gravitation wave occurs in the near zone. The propagation and detection of gravitational wave happens in the far zone, where the Riesz effect becomes valuable. Once emitted, the gravitational wave diffuses and it amplitude sharply reduces. The amplitude of the Riesz wave reduces as $r^{-2}$, being much weaker then D'Alembert wave with the amplitude decay of the order $r^{-1}$. The recent data form LIGO in situ experiment claims no detected signal of the expected amplitudes [29].

## Conclusions

The above results may be summed up as follows. The idea of new purely speculative hypothesis consists of the introduction of isotropic partial field equation in form of quadratic polynomial. The linear term is usually omitted due to lack of isotropic partial differential operators of first order. There exist, however, the isotropic *pseudodifferential* operators of the first order. This purely technical generalization leaves sufficient room for the extension of the basic equations and explains some of the "dark mass" phenomena.

The relativistic invariant, scale-variant equations of post-Newtonian gravity were derived. For derivation of the governing equations we use the variational principles with the partial fractional differentials. The operator for static gravitation potential is the weighted sum of Laplace operator and elliptical Riesz potential operator. In non-stationary case, the modified wave operator constitutes as sum of D'Alembert wave operator and Riesz hyperbolical operator. The corresponding spherically symmetric and Lorenz invariant Hamiltonian contains fractional derivatives. The governing equations are presumed to be generally covariant, relativistic invariant and spatially isotropic. The solitary constant of the theory is the characteristic length $\lambda$. Gravitational field, matches to (i) the weak Newtonian limit on the moderate scales, (ii) delivers a oscillating Post- Newtonian potential on characteristic galactic scales. This feature explains many observed properties of galaxies without the necessity of non-baryonic dark matter.

On the interstellar scales, the limit for the proposed model corresponds those of the classical General Relativity. The Riesz potential perturbation declines rapidly than Newtonian potential on the short scales, such that it has insignificant exposure on the processes with short duration, with time and space sizes of Solar system.

At the certain galactic scales Riesz gravitational acceleration expected to be much higher then Newtonian. This remarkable feature could explain many observed properties of galaxies without the necessity of non-baryonic dark matter. The proposed model eliminates in many cases the need for the dark, non-observable mass. The hypothesis explains maximal sizes of galaxies and predicts the regions of stability. The dwarf members of Local Group fall within the regions of stability.

The Green function for the wave equation with Riesz term and the representation of the solution for fractional wave equation are found in form of retarded potentials. The solutions for the Riesz wave equation and classical wave equation are clearly distinctive in an important sense. The Riesz wave demonstrates the space diffusion of gravitational wave at the scales of metric constant. The diffusion leads to the blur of the peak and disruption of the sharp wave front. This contrasts with the solution of D'Alembert classical wave equation, which obeys the Huygens principle and does not diffuse. Yet the proposed model is of pure speculative nature, it delivers an unusual explanation of some really observed phenomena.

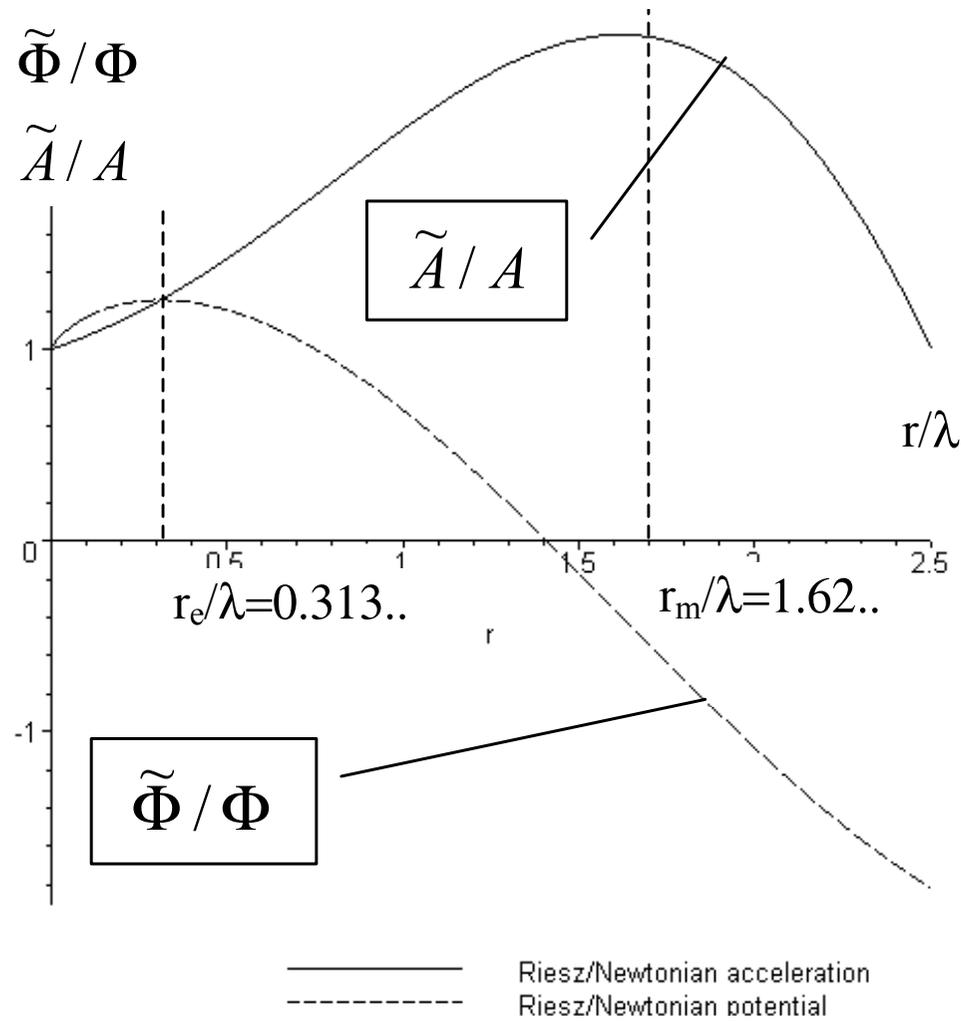

Fig. 1 Riesz and Newtonian potentials and accelerations on the short scales

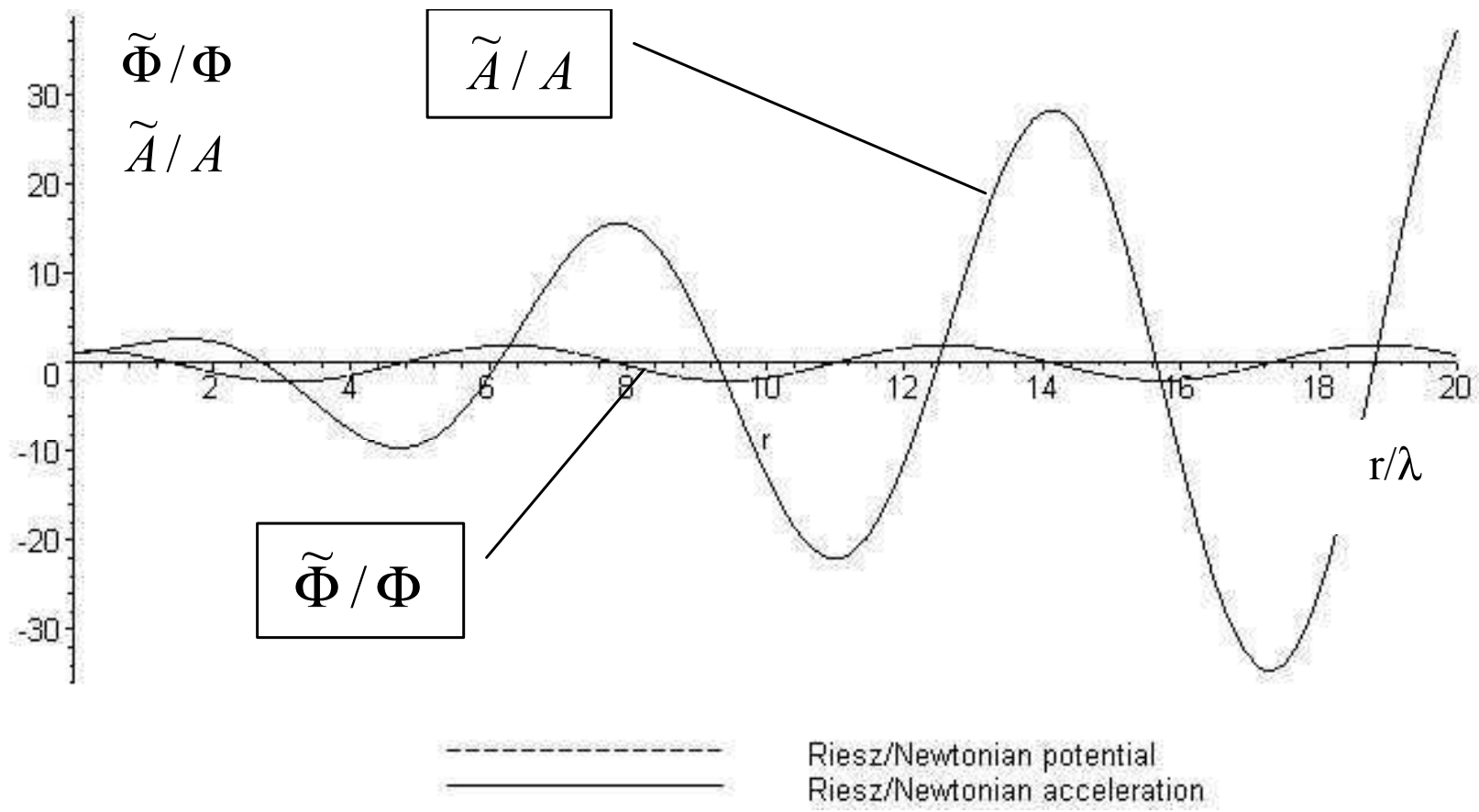

Fig. 2 Ratio of Riesz/Newtonian potentials and accelerations for large values of $r/\lambda$

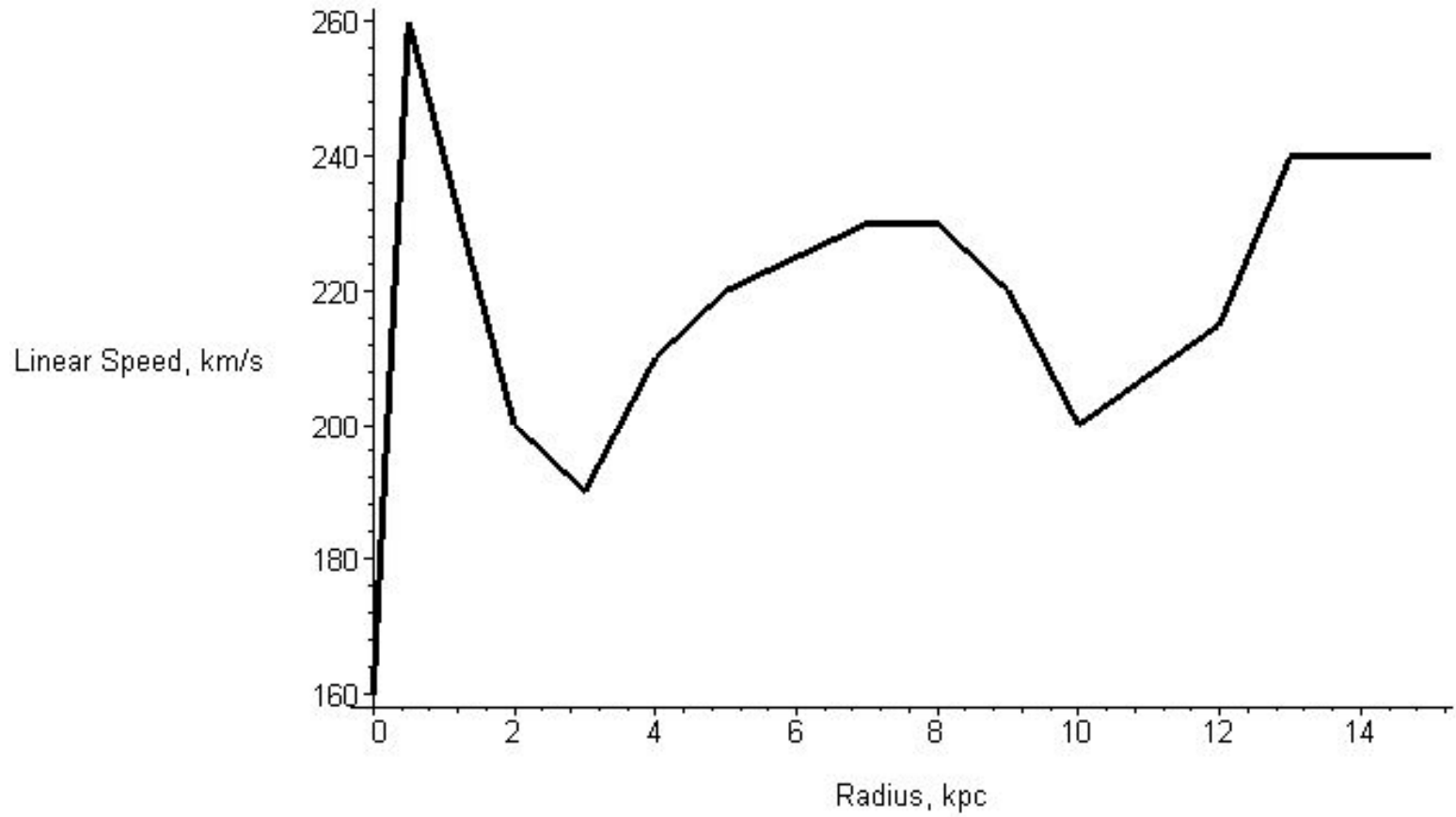

Fig. 3 Rotation curve of Milky Way as a function of distance to Galaxy center

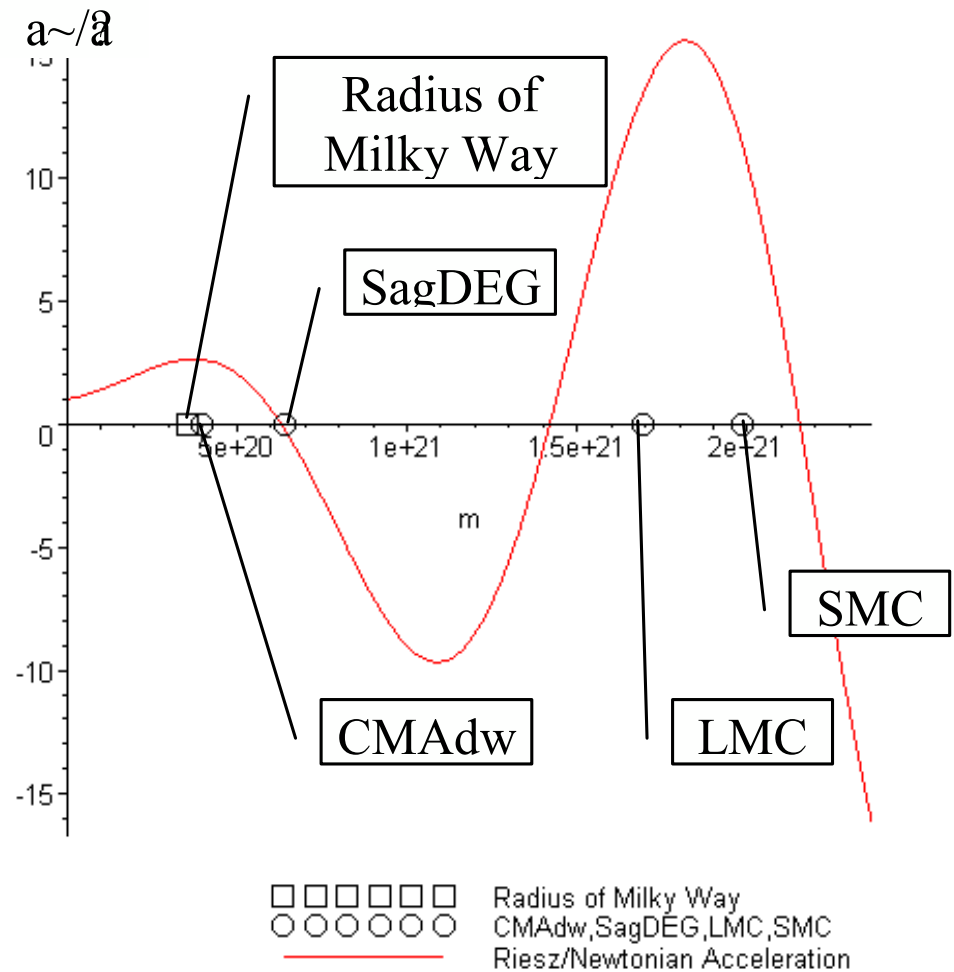

Fig. 4 Regions of stability and the distances from nearest Local Group Galaxies to the center of Milky Way

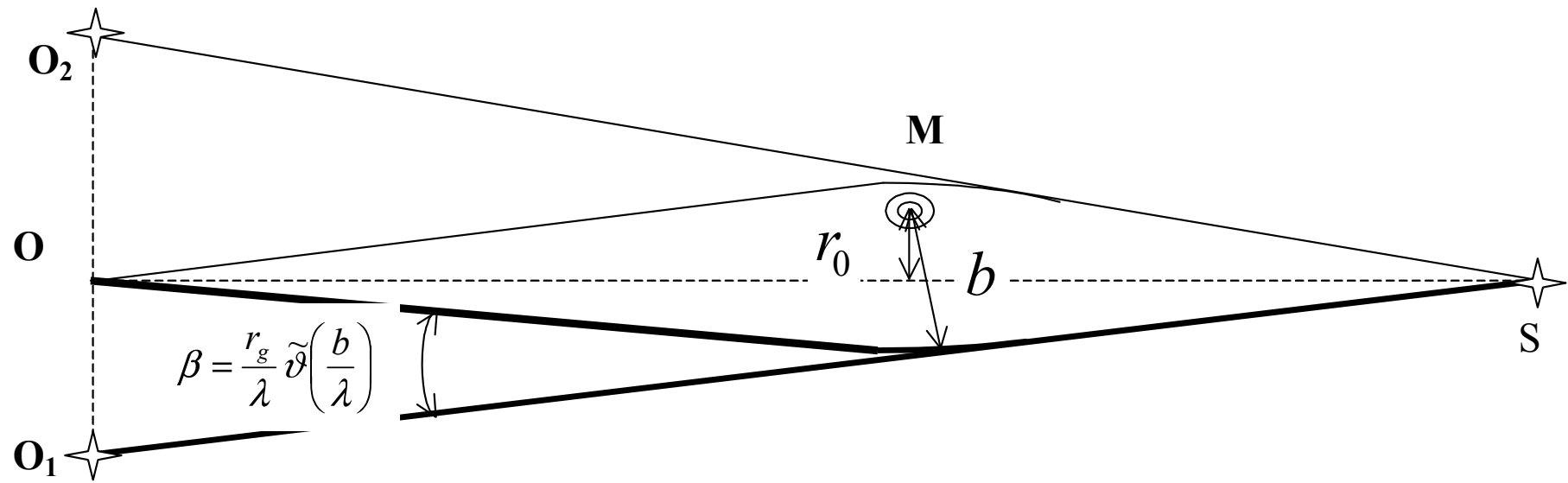

Fig. 5 Geometry of gravitational lens

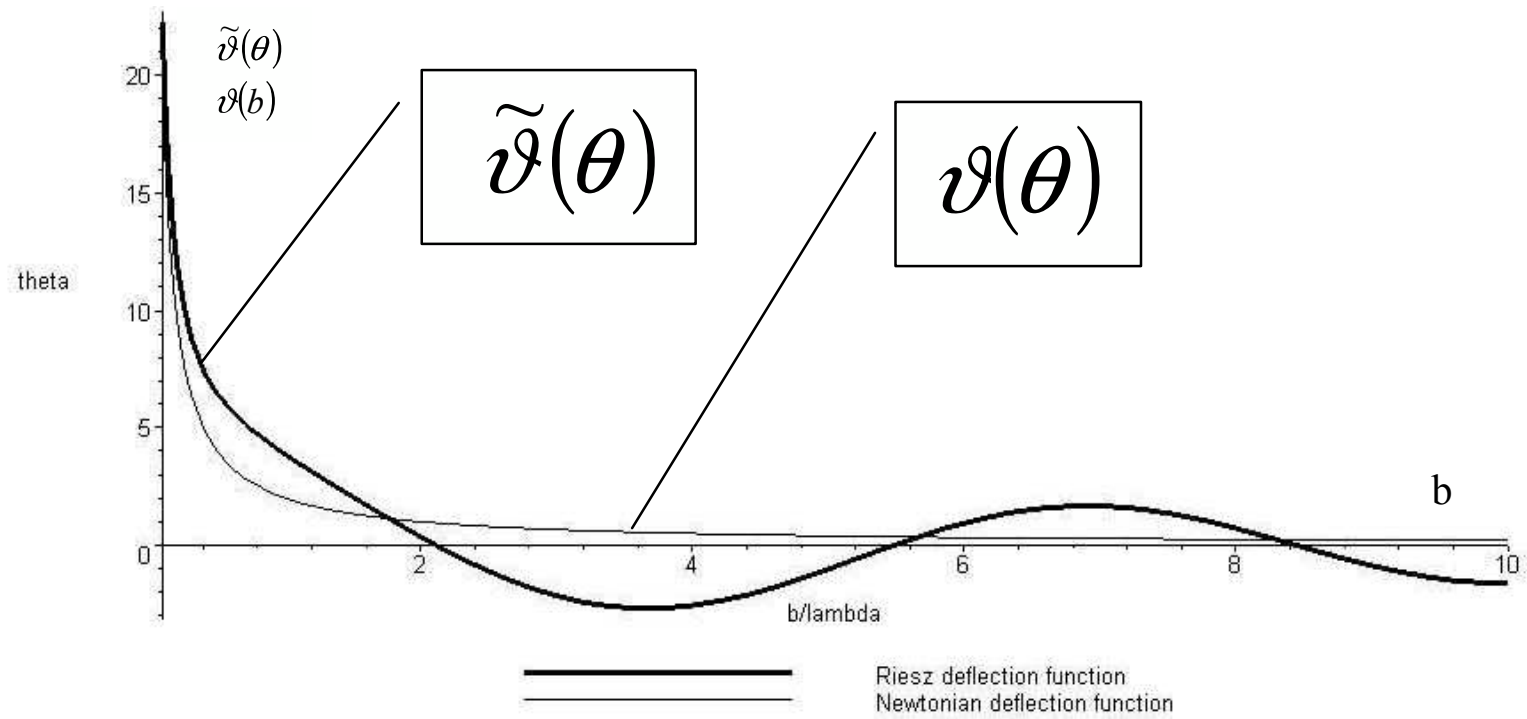

Fig. 6 Deflection functions for Riesz and Newtonian potentials as functions of impact parameter

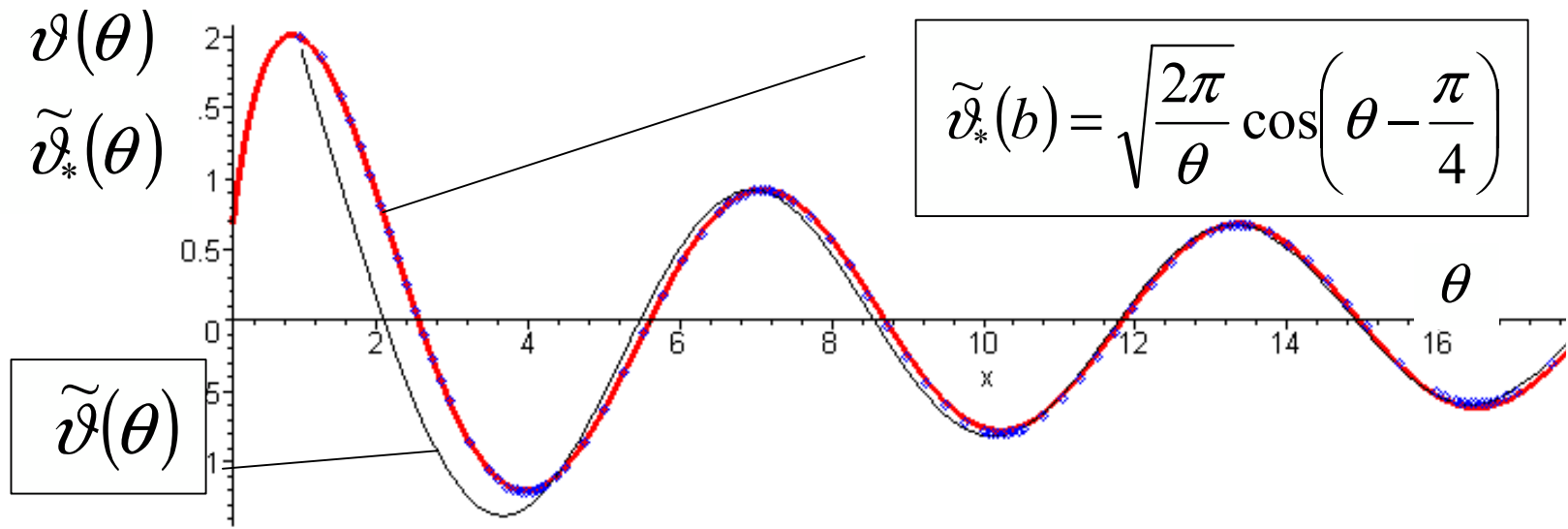

Fig. 7 Deflection functions for Riesz and Newtonian potentials as functions of impact parameter